\documentclass[pre,twocolumn,a4paper,10pt,amssymb,amsmath]{revtex4}
\usepackage[dvips]{graphicx}

\newtheorem{conjecture}{Conjecture}

\newcommand{\var}[1]{\mathrm{var}\left( #1 \right)}
\newcommand{\avg}[1]{\left< #1 \right>}
\newcommand{\fZ}{\mathcal{Z}}

\newcommand{\fU}{\mathcal{U}}
\newcommand{\fC}{\mathcal{C}}
\begin{document}

\title{Non-vanishing boundary effects and quasi-first order phase
  transitions in high dimensional Ising models}

\date{\today}
\author{P. H. Lundow} 
\email{phl@kth.se} 
\affiliation{
  Condensed Matter Theory, Department of Theoretical Physics,\\
  AlbaNova University Center, KTH, SE-106 91 Stockholm, Sweden 
}
\author{K. Markstr\"om}
\email{klas.markstrom@math.umu.se}
\affiliation{
  Department of mathematics and mathematical statistics,
  Ume\aa{} University, SE-901 87 Ume\aa, Sweden
}

\begin{abstract}
  In order to gain a better understanding of the Ising model in higher
  dimensions we have made a comparative study of how the boundary, open
  versus cyclic, of a $d$-dimensional simple lattice, for
  $d=1,\ldots,5$, affects the behaviour of the specific heat $\fC$ and
  its microcanonical relative, the entropy derivative $-\partial S /
  \partial U$.
  
  In dimensions 4 and 5 the boundary has a strong effect on the
  critical region of the model and for cyclic boundaries in dimension
  5 we find that the model displays a quasi first order phase
  transition with a bimodal energy distribution. The latent heat
  decreases with increasing systems size but for all system sizes used
  in earlier papers the effect is clearly visible once a wide enough
  range of values for $K$ is considered.
  
  Relations to recent rigorous results for high dimensional
  percolation and previous debates on simulation of Ising models and
  gauge fields are discussed.
  
\end{abstract}

\maketitle
\section{Introduction}
The Ising model on 2 and 3-dimensional lattices is probably the most
studied model in the theory of phase transitions from the condensed
matter point of view. However, since at least the early 90's higher
dimensional lattices have become increasingly more important due to
their connection to gauge field theory and particle physics. In
particular the Ising model on the 4- and 5-dimensional cubic lattices
have been studied both theoretically and via simulation.  In these
dimensions it is known that in the thermodynamical limit
\cite{aizenman:82,sokal:79} the model follows the mean field critical
exponents, but the finite size behaviour of the model has been
hotly debated \cite{brezin:85, mon:90, mon:96, PhysRevB.31.1498,
  Stevenson:05, BN:06, cd:98a, CD:98b, CD:00, CCC:98, LBB:99, BL:97,
  luijten:99}. Part of the importance of these debates comes from the
fact that it directly impinges on the methods used to derive other
quantities from the finite size data, whose values are not already
known in the limit.  These methods e.g. affect what has been done in
field theoretical studies of the Higgs particle mass as well as other
properties studied using lattice quantum chromodynamics.

In the work presented here we set out to make a systematic comparison
of the finite size behaviour of the Ising model on lattices with open
and cyclic boundary conditions. We include both the well studied 2-
and 3-dimensional lattices and the earlier debated lattices in
dimension 4 and 5. To our knowledge the previous studies in higher
dimensions have focused exclusively on lattices with cyclic boundary
conditions, but as recent rigorous results on percolation
\cite{aizenman:97, HH:07} has shown, for this, simpler, model
there are non-vanishing boundary effects in higher dimension. We
believe that there is a risk in focusing only the cyclic case.

As our sampled data reveals, there is a striking change in how
the boundary conditions affect the finite size behaviour as we pass
the critical dimension $D=4$.  For $D=4$ the finite size effects are
more visible in the microcanonical ensemble than in the
canonical. This is due to the smoothing effect of fluctuations in the
energy of the system. Here the data, as previously shown in
\cite{4dart}, favours the conclusion that the specific heat at $K_c$ is
bounded, which is consistent with the rigorous result $\alpha=0$ but
in conflict with predictions of a weak divergence by renormalization
theory \cite{ptcp}.  We also find that for a certain range of small
sizes and open boundary conditions the specific heat has \emph{two}
maxima close to the critical point, one of which disappears as the
system size increases. In the microcanonical ensemble this effect
remains visible even for the largest systems we have studied (L=40).
For cyclic boundary conditions not such effects are visible in either
ensemble.

For $D=5$, where the specific heat is rigorously known to be bounded
\cite{sokal:79}, the model display more striking differences between
the two boundary conditions. As for the 4-dimensional case there is a
system size range with two maxima in the specific heat for free
boundary conditions, but not for the cyclic case. This effect is again
visible for even larger size in the microcanonical ensemble, because
the smoothing effect of energy fluctuations is not present here. There
is also a larger gap between the effective critical points for each
size, e.g. the values of $K$ at which the model has the maximum
specific heat, but this is to be expected due to the relatively large
boundary size.  However, the most striking difference is that for
cyclic boundary conditions the model displays a quasi-first order
phase transition. For $K$ close to $K_c$ the energy distribution
becomes bimodal, just as for a high-$q$ Potts model. As the system size
increases the gap between the two maximima scales as $L^{-5/2}$,
thereby giving the model an effective latent heat for finite size, but
not in the thermodynamic limit. Due to this bimodal behaviour
simulations become sensitive to small changes in $K$, as
this makes the model favour one of the two maxima and thereby a
different internal energy. In fact, some of the disagreements between
different simulations and data mentioned in \cite{brezin:85, mon:90,
  mon:96, PhysRevB.31.1498, Stevenson:05, BN:06, cd:98a, CD:98b,
  CD:00, CCC:98, LBB:99, BL:97, luijten:99} might be due to this
overlooked effect.

\section{Notation}
We will investigate the behaviour of two types of lattice graphs, the
cartesian graph product of $d$ cycles, i.e. $C_L \times \cdots \times
C_L$, and the product of $d$ paths $P_L\times \cdots \times P_L$ for
$d=2,3,4,5$. Both types of graphs have $n=L^d$ vertices, the product
of cycles has $d\,L^d$ edges and the product of paths has
$d\,L^{d-1}\,(L-1)$ edges. They differ only at their boundaries
where the cycle product has cyclic boundary conditions
while the path product has open boundary conditions.

We define the energy of a state $\sigma=(\sigma_1,\ldots,\sigma_n)$,
with $\sigma_i=\pm 1$, as $E(\sigma) = \sum_{\{i,j\}}
\sigma_i\,\sigma_j$ (summation over the edges $\{i,j\}$), and the
magnetisation as $M(\sigma)=\sum_i\sigma_i$ (summation over the
vertices).

We look into two classes of quantities. The first of these, the
combinatorial quantities from the microcanonical ensemble, provide an
extremely detailed picture of the inner workings of the model, and
they all depend on the energy $U=E/n$. The coupling $K$, defined as
$K(U) = -\partial S / \partial U$, is of special interest here. Here
$S(U) = (1/n)\,\ln a(E)$ where $a(E)$ denotes the number of states
$\sigma$ at energy $E$.  It is described in great detail in
\cite{sampart} how to obtain these from sampled data. 

The second class of quantities, the canonical (or physical) ones, are
the cumulants, i.e. derivatives of $\ln\fZ$ where $\fZ$ is the
partition function.  How to obtain these quantities from the coupling
function $K(U)$ is described in \cite{reconart}. We will refer to them
in their dimensionless forms.  We define the internal energy
as $\fU(K)=\avg{E}/n$, the specific heat as $\fC(K)=\var{E}/n$
and the susceptibility as $\chi(K) = \var{M}/n=\avg{M^2}/n$.
For each coupling $K$ we receive a distribution of energies having
density function $\Pr{E}$ and again we refer the reader to
\cite{reconart} for how to obtain this distribution from the
$K(U)$-function.

Occasionally we will, when necessary, subscript functions with the
linear order $L$ of the lattice, as e.g. $\fC_L(K)$.  We denote by
$K^*$ the coupling where $\fC_L$ is at its maximum.  We denote by
$U^*$ an energy where $\partial K/\partial U$ has a local minimum and
we use $U^+$ for a second minimum where $U^*<U^+$.

\section{2D-lattices}
\begin{figure}[!hbt]
  \begin{center}
    \includegraphics[width=0.49\textwidth]{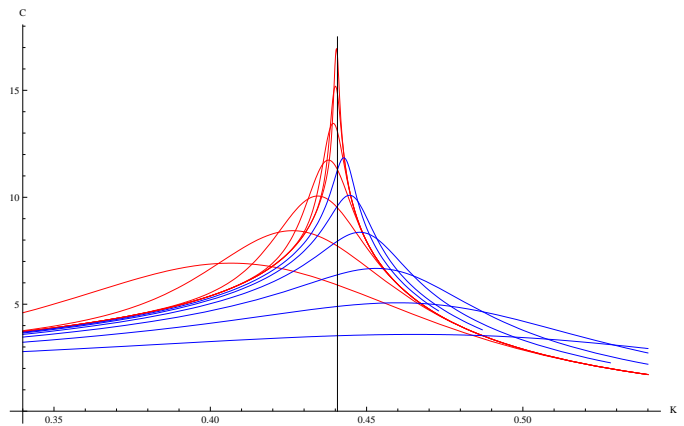}
  \end{center}
  \caption{(Colour online) Specific heat $\fC(K)$ for 2D cycle
    products (red) of linear order $L=8$, $16$, $32$, $64$, $128$,
    $256$, $512$ and for path products (blue) of linear order $L=8$,
    $16$, $32$, $64$, $128$, $256$. Black vertical line at
    $K_c=0.44068\ldots$.}
  \label{fig:c2d}
\end{figure}

For 2-dimensional lattices there is no qualitative difference between
the cycle and the path product for the specific heat's behaviour. In
Figure~\ref{fig:c2d} we plot the specific heat for both lattice types
for a range of linear orders.  Though the specific heat is somewhat
smaller for the path products, not so strange considering that these
graphs have fewer edges, the only substantial feature of the path
products is that the peak is always located to the right of $K_c$
rather than to the left as it is for cycle products.  We should here
mention that for cycle products with $L\le 320$ we rely on exact data
computed in \cite{exactart} and for $L=512$ on sampled data. For path
products we have exact data only for $L\le 16$, see \cite{lundow:99b},
and sampled data otherwise.

\begin{figure}[!hbt]
  \begin{center}
    \includegraphics[width=0.49\textwidth]{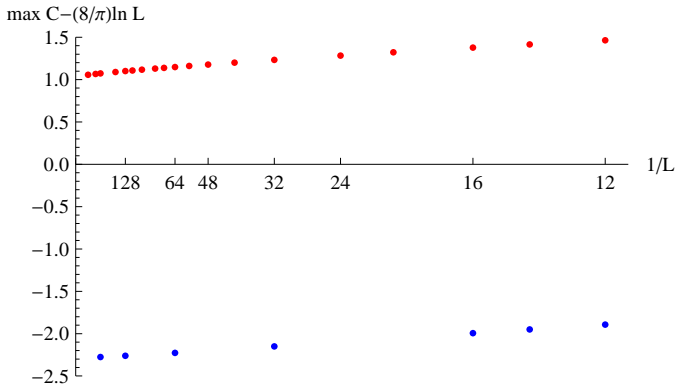}
  \end{center}
  \caption{(Colour online) $\max \fC_L - (8/\pi)\,\ln L$ vs $1/L$ for
    2D cycle products (red points) of linear order $L=12$, $14$, $16$,
    $20$, $24$, $32$, $40$, $48$, $56$, $64$, $72$, $80$, $96$, $112$,
    $128$, $160$, $256$, $320$, $512$ and for path products (blue
    points) of linear order $L=12$, $14$, $16$, $32$, $64$, $128$,
    $256$.}
  \label{fig:cmaxval2d}
\end{figure}

Let us compare how the maximum specific heat grows. We know that for
square cycle products the maximum grows as $1.03684 + (8/\pi)\,\ln L$,
and we extract this from \cite{fisher:69}. Note that for
$C_{\infty}\times C_L$ the additive constant is instead $0.967550$,
see \cite{onsager:44} for an exact expression. In
Figure~\ref{fig:cmaxval2d} we show $\max \fC_L - (8/\pi)\,\ln L$
versus $1/L$ for both types of lattices. Though we do not know what
the limit additive constant will be for path products we can estimate
it to be $-2.30$.

\begin{figure}[!hbt]
  \begin{center}
    \includegraphics[width=0.49\textwidth]{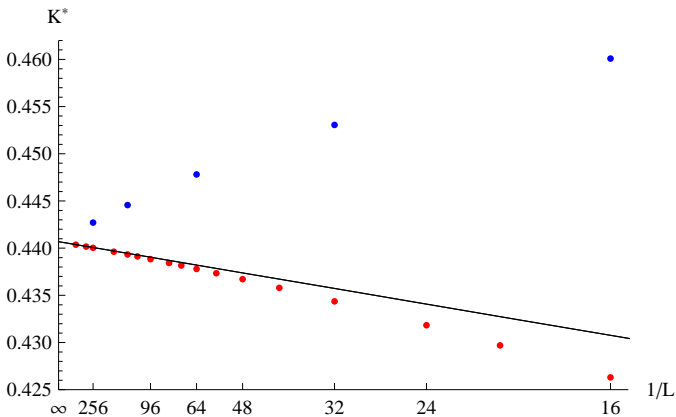}
  \end{center}
  \caption{(Colour online) Location $K^*$ of maximum specific heat vs
    $1/L$ for 2D cycle products (red points) of linear order $L=16$,
    $20$, $24$, $32$, $40$, $48$, $56$, $64$, $72$, $80$, $96$, $112$,
    $128$, $160$, $256$, $320$, $512$ and for path products (blue
    points) of linear order $L=16$, $32$, $64$, $128$, $256$. The
    black asymptote is $K_c - 0.15878/L$.}
  \label{fig:cmaxloc2d}
\end{figure}

The locations of the maxima are plotted in Figure~\ref{fig:cmaxloc2d}.
Note that the asymptotic behaviour is already known to be $K_c-K^*\sim
0.15878/L$, see~\cite{fisher:69}, for the cycle products. For path
products there are obviously some higher order correction terms at
work as well.

\begin{figure}[!hbt]
  \begin{center}
    \includegraphics[width=0.49\textwidth]{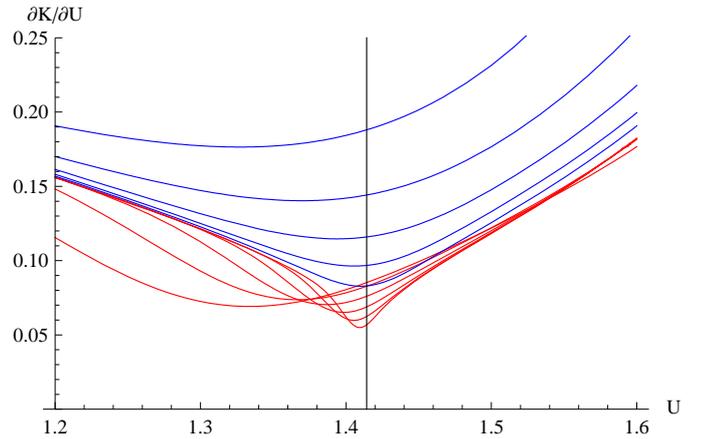}
  \end{center}
  \caption{(Colour online) $\partial K/\partial U$ for 2D cycle
    products (red) of linear order $L=16$, $32$, $64$, $128$, $256$,
    $512$ and for path products (blue) of linear order $L=16$, $32$,
    $64$, $128$, $256$. Black line at $U_c=\sqrt{2}$.}
  \label{fig:kd2d}
\end{figure}

Looking from a microcanonical perspective and plotting the derivative
$\partial K/\partial U$, which is effectively a microcanonical inverse
of the specific heat, we see a more or less similar behaviour between
the lattice types in Figure~\ref{fig:kd2d}. This time the lattice
types agree on which side of the asymptotic location the local
minimum should be.

\begin{figure}[!hbt]
  \begin{center}
    \includegraphics[width=0.49\textwidth]{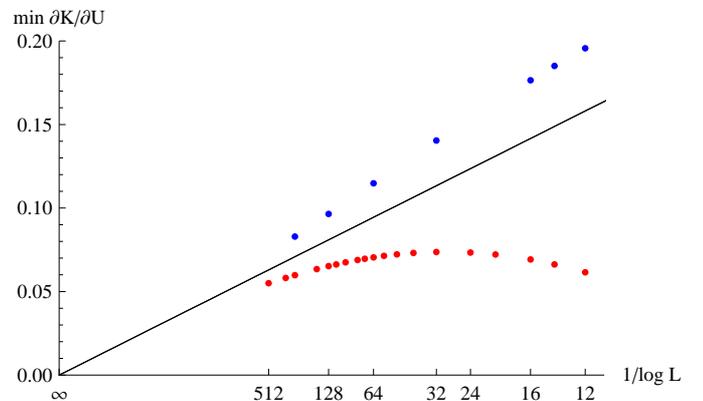}
  \end{center}
  \caption{(Colour online) $\min\partial K/\partial U$ for 2D cycle
    products (red points) of linear order $L=12$, $14$, $16$, $20$,
    $24$, $32$, $40$, $48$, $56$, $64$, $72$, $80$, $96$, $112$,
    $128$, $160$, $256$, $320$, $512$ and for path products (blue
    points) of linear order $L=12$, $14$, $16$, $32$, $64$, $128$,
    $256$. Black line is the asymptote $\pi/(8\,\ln L)$.}
  \label{fig:kdminval2d}
\end{figure}

For these lattices we can actually give the asymptotical behaviour for
how the minimum of $\partial K/\partial U$ should approach $0$.  Since
the maximum specific heat grows as $(8/\pi)\,\ln L$, and we excpect no
asymyptotic difference here between the lattice types, the minimum
$\partial K/\partial U$ must vanish at exactly the inverse rate
$\pi/(8\,\ln L)$. This is shown in Figure~\ref{fig:kdminval2d}. Note
that the path product lattices approach the asymptote from above and
the cycle products from below.

\section{3D-lattices}
For 3-dimensional lattices we see no major difference in behaviour
between the two lattices types either. In three dimensions we do not 
 an exact solution to rely on, so we can not discern as many details
as for 
 2D-lattices.  For cycle products we have sampled data
for $L=6,8,12,16,32,64,128,256,512$, as used in \cite{cubeart} while
for path products we have sampled data only for smaller lattices;
$L=6,8,12,16,24,32,46,64,96,128$. For both lattice types we use exact
data for $L=4$.

\begin{figure}[!hbt]
  \begin{center}
    \includegraphics[width=0.49\textwidth]{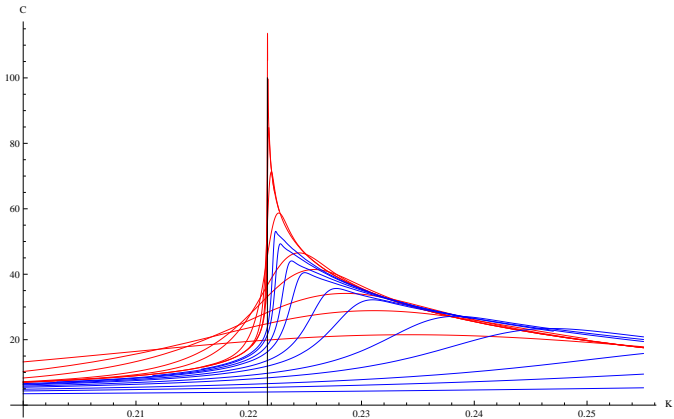}
  \end{center}
  \caption{(Colour online) Specific heat $\fC(K)$ for 3D cycle
    products (red) of linear order $L=4$, $6$, $8$, $12$, $16$, $32$,
    $64$, $128$, $256$, $512$ and for path products (blue) of linear
    order $L=4$, $6$, $8$, $12$, $16$, $24$, $32$, $48$, $64$, $96$,
    $128$. Black vertical line at $K_c\approx 0.2216546$.}
  \label{fig:c3d}
\end{figure}

\begin{figure}[!hbt]
  \begin{center}
    \includegraphics[width=0.49\textwidth]{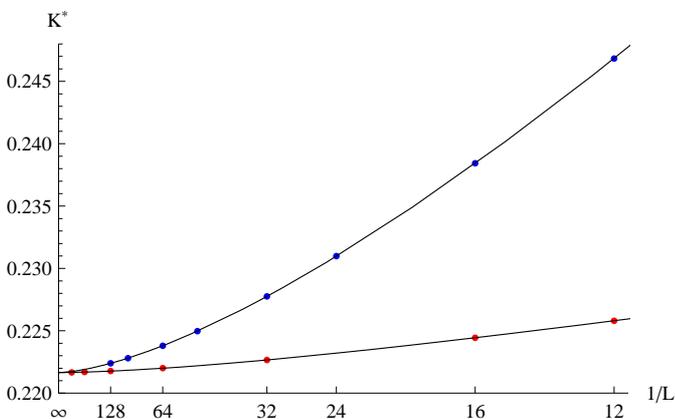}
  \end{center}
  \caption{(Colour online) Location $K^*$ of maximum specific heat for
    3D cycle products (red) of linear order $L=4$, $6$, $8$, $12$,
    $16$, $32$, $64$, $128$, $256$, $512$ and for path products (blue)
    of linear order $L=4$, $6$, $8$, $12$, $16$, $24$, $32$, $48$,
    $64$, $96$, $128$. See text for parameters of fitted curves.}
  \label{fig:cmaxloc3d}
\end{figure}

Again let us compare the lattice types. The specific heat is shown in
Figure~\ref{fig:c3d}. As in the 2D-case the path products' maximum
specific heat is smaller. 
The location of the maximum approach $K_c$ from above for both lattice
types, see Figure~\ref{fig:cmaxloc3d}. However, for path products this
maximum is located much farther away from $K_c$. The black curves in
the plot were obtained from a best fit of all available data points to
the simple scaling formula
\begin{equation}\label{scale}
  K^* = c_0 + c_1\,L^{-\lambda_1} + c_2\,L^{-\lambda_2}
\end{equation}
where we pre-ordained $c_0=0.2216546$ as found in \cite{cubeart}.  For
the cycle products we received $c_1=0.3775$, $\lambda_1=1.654$,
$c_2=-0.6169$ and $\lambda_2=2.297$.  For the path products this gave
$c_1=2.292$, $\lambda_1=1.617$, $c_2=-2.101$ and
$\lambda_2=1.961$. The fitted curves agree beautifully with the points
for all $L$ and they are included in the plot.

3D
$16$,
$64$,

\begin{figure}[!hbt]
  \begin{center}
    \includegraphics[width=0.49\textwidth]{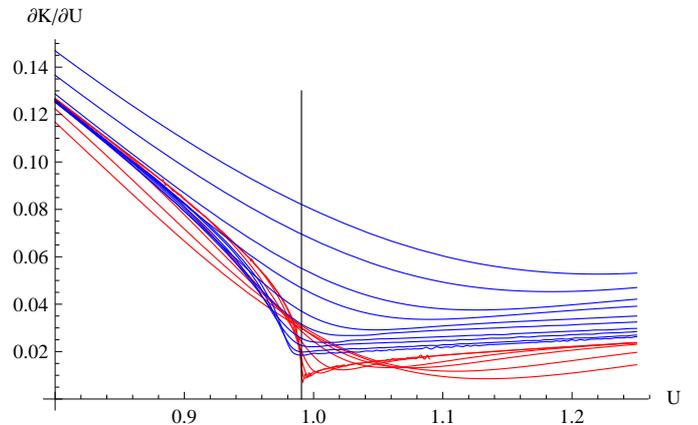}
  \end{center}
  \caption{(Colour online) $\partial K/\partial U$ for 3D cycle
    products (red) of linear order $L=6$, $8$, $12$, $16$, $32$, $64$,
    $128$, $256$, $512$ and for path products (blue) of linear order
    $L=6$, $8$, $12$, $16$, $24$, $32$, $48$, $64$, $96$, $128$. Black
    line at $U_c\approx 0.99063$.}
  \label{fig:kd3d}
\end{figure}

\begin{figure}[!hbt]
  \begin{center}
    \includegraphics[width=0.49\textwidth]{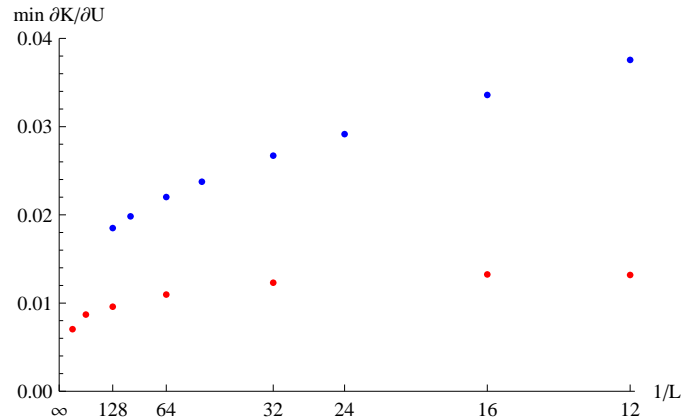}
  \end{center}
  \caption{(Colour online) Minimum $\partial K/\partial U$ for 3D
    cycle products (red) of linear order $L=6$, $8$, $12$, $16$, $32$,
    $64$, $128$, $256$, $512$ and for path products (blue) of linear
    order $L=6$, $8$, $12$, $16$, $24$, $32$, $48$, $64$, $96$,
    $128$. }
  \label{fig:kdminval3d}
\end{figure}

On the microcanonical side we note that the derivative $\partial
K/\partial U$ is much smaller for cycle products than for path
products, see Figure~\ref{fig:kd3d}. In Figure~\ref{fig:kdminval3d} we
show the minimum values for both lattice types. Note that for cycle
products this minimum is actually increasing for $L<16$, whereas for
path products they are consistently decreasing. This decrement speeds
up dramatically, as it should, for $L>64$ as shown in the figure.

\section{4D-lattices}
In \cite{4dart} the authors studied the Ising model on cycle products
and, unexpectedly, found that the specific heat seemed to be bounded
at $K_c$, a result which is compatible with the rigorous bounds but in
disagreement with non-rigorous renormalization predictions. As we
shall now see, unlike in the lower dimensional cases, in 4 dimensions
there are also distinct differences in the finite size behaviour of
the cycle and path products. In short, the specific heat of the path
and cycle products differ somewhat in their behaviour, however on the
microcanonical side larger differences are clearly visible. The
minimum of $\partial K/ \partial U$ at $U^*$ is much higher for path
products and is decreasing with $L$, compared to the cycle products
where it is actually increasing. Also, for $L\ge 12$ there is a second
local minimum in $\partial K/ \partial U$ at $U^+$, where
$U^*<U^+$. This minimum becomes a global minimum for $L\ge 24$. We
will now proceed as for the other lattices.

\begin{figure}[!hbt]
  \begin{center}
    \includegraphics[width=0.49\textwidth]{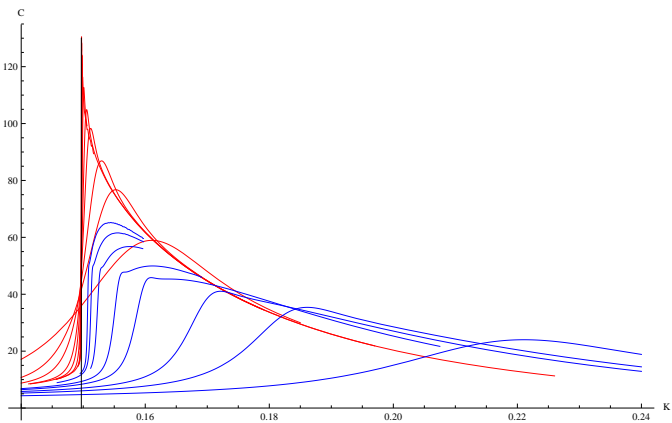}
  \end{center}
  \caption{(Colour online) Specific heat $\fC(K)$ for 4D cycle
    products (red) of linear order $L=4$, $6$, $8$, $12$, $16$, $24$,
    $32$, $40$, $48$, $64$, $80$ and for path products (blue) of
    linear order $L=4$, $6$, $8$, $12$, $16$, $24$, $32$, $40$. Black
    vertical line at $K_c\approx 0.1496947$.}
  \label{fig:c4d}
\end{figure}

\begin{figure}[!hbt]
  \begin{center}
    \includegraphics[width=0.49\textwidth]{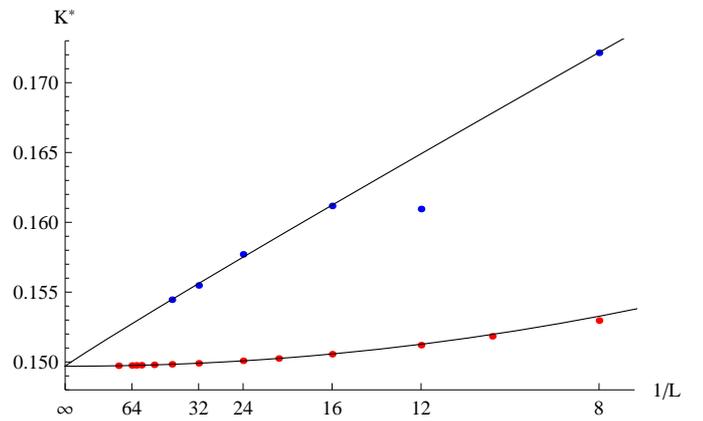}
  \end{center}
  \caption{(Colour online) Location $K^*$ of the maximum specific heat
    $\fC(K)$ for 4D cycle products (red) of linear order $L=4$,
    $6$, $8$, $10$, $12$, $16$, $20$, $24$, $32$, $40$, $48$, $56$,
    $60$, $64$, $80$ and for path products (blue) of linear order
    $L=4$, $6$, $8$, $12$, $16$, $24$, $32$, $40$. }
  \label{fig:cmaxloc4d}
\end{figure}

The specific heat is shown in Figure~\ref{fig:c4d}.  Note especially
the peculiar behaviour of $\fC$ near $K^*$ for the path products.
This is a phenomenon due to the extra local minimum at $U^+$ in
$\partial K/ \partial U$. Due to this peculiarity the location of the
global maximum will display an irregular behaviour, depicted in
Figure~\ref{fig:cmaxloc4d}.

In Figure~\ref{fig:cmaxloc4d} we have fitted curves to $K^*$ that are
forced to have the same $c_0 = K_c = 0.1496947$, obtained in
\cite{4dart}, and obviously there is no conflict in this. However, the
correct asymptotic behaviour for the path products is only obtained
from $L\ge 16$, for our choice of lattice sizes, leaving us only four
points to fit our curve to. Thus we set $c_2=0$ in both cases and
obtain $c_1=0.243$ and $\lambda_1=2.03$ for the cycle products. For
the path products we obtain $c_1=0.166$ and
$\lambda_1=0.962$. Strikingly, the exponents differ by about a factor
two.

\begin{figure}[!hbt]
  \begin{center}
    \includegraphics[width=0.49\textwidth]{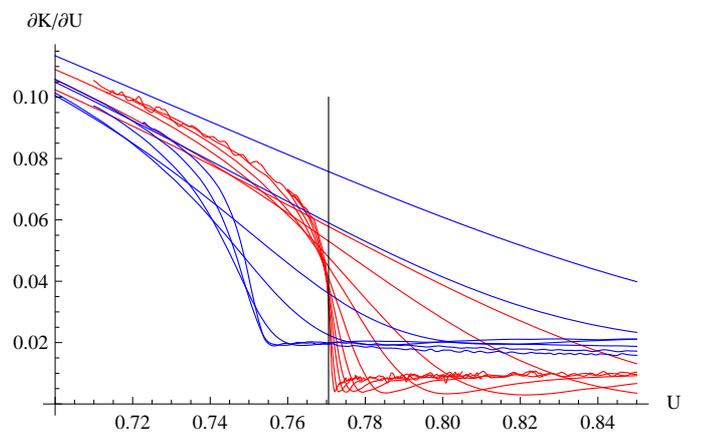}
  \end{center}
  \caption{(Colour online)  $\partial K/ \partial U$
    for 4D cycle products (red) of linear order
    $L=4$, $6$, $8$, $12$, $16$, $24$, $32$, $40$, $48$,
    $64$, $80$ and for path products (blue) of linear
    order $L=4$, $6$, $8$, $12$, $16$, $24$, $32$, $40$. Black
    line at $U_c\approx 0.770527$.}
  \label{fig:kd4d}
\end{figure}

\begin{figure}[!hbt]
  \begin{center}
    \includegraphics[width=0.49\textwidth]{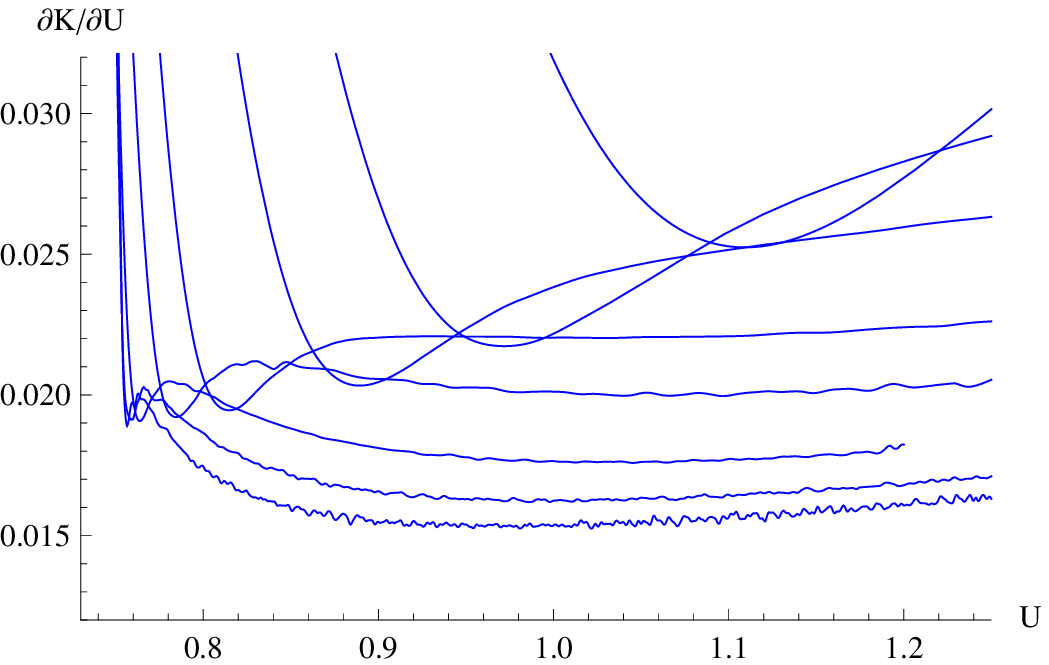}
  \end{center}
  \caption{(Colour online) $\partial K/ \partial U$ for 4D path
    products of linear order $L=4$, $6$, $8$, $12$, $16$, $24$, $32$,
    $40$.}
  \label{fig:kdp4d}
\end{figure}

\begin{figure}[!hbt]
  \begin{center}
    \includegraphics[width=0.49\textwidth]{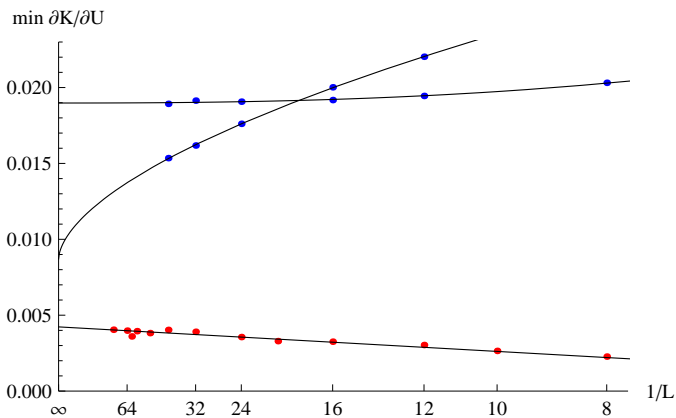}
  \end{center}
  \caption{(Colour online) Minimum value of $\partial K/ \partial U$
    for 4D cycle products (red, converging to $0.004$) of linear order
    $L=4$, $6$, $8$, $10$, $12$, $16$, $20$, $24$, $32$, $40$, $48$,
    $56$, $60$, $64$, $80$ and for path products (blue) of linear
    order $L=4$, $6$, $8$, $12$, $16$, $24$, $32$, $40$. The blue
    points converging to $0.019$ are the minima located at $U^*$ and
    those converging to $0.009$ are located at $U^+$.}
  \label{fig:kdminval4d}
\end{figure}

Let us now turn to the microcanonical side. In Figure~\ref{fig:kd4d}
we show $\partial K/ \partial U$ for both lattice types, near $U_c$,
where each curve has a local minimum. Outside the picture there also
materialises, for path products with $L\ge 12$, a second local
minimum, see Figure~\ref{fig:kdp4d}. This minimum is not as sharp as
the other minimum, thus making it difficult to pin-point it to a high
accuracy. Moreover, this minimum becomes a global minimum for $L\ge
24$. This phenomenon is shown in Figure~\ref{fig:kdminval4d}.

Here the picture gets more complicated. First, as was demonstrated in
\cite{4dart}, the values of the local (and only) minima for the cycle
products are increasing to about $0.0042$, thus leading to an
asymptotically finite specific heat. As discussed in discussed in
\cite{4dart} this is in disagreement with a prediction from
renormalization theory, which predicts a logarithmic singularity in
the specific heat for $D=4$. It is rigorously known \cite{sokal:79}
that the ciritcal exponent $\alpha=0$ for $D=4$. In \cite{4dart} data
from both a single spin Metropolis and a Wolff-cluster algorithm
\cite{wolff:89} were compared to rule out methodological errors and
the conclusion was that either systems as large as $L=80$ are still
dominated by finite size effects or the renormalization prediction
fails for the $D=4$ case, which is known to be the critical dimension
for the nearest neighbour Ising model.

Second, the local minimum nearest $U_c$ for the path products seems to
approach a different and even greater value of roughly $0.019$. Third,
the other local minimum for the path products, also becoming a global
minimum, could approach yet another value of about $0.009$ in between
the other two.  Forcing this minimum to approach the same minimum as
for the cycle products seems out of the question with our current set
of data though.

In Figure~\ref{fig:kdminval4d} the fitted curves a la (\ref{scale})
use $c_2=0$ and we received $c_0=0.0042$, $c_1=-0.016$, when setting
$\lambda_1=1$ for the cycle products, see a discussion on this in
\cite{4dart}. For the first set of local minima for the path products,
the ones nearest $U_c$ we received $c_0=0.0190$, $c_1=0.223$,
$\lambda_1=2.47$. The second set for the path products, farther away
from $U_c$, gave us $c_0=0.00870$, $c_1=0.0568$, $\lambda_1=0.583$
when the curve was fitted to all data points.

\begin{figure}[!hbt]
  \begin{center}
    \includegraphics[width=0.49\textwidth]{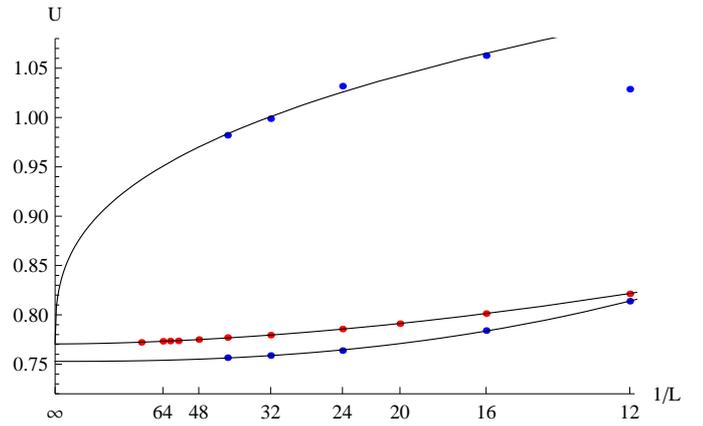}
  \end{center}
  \caption{(Colour online) Location $U^*$ of the minimum value of
    $\partial K/ \partial U$ for 4D cycle products (red points) of
    linear order $L=4$, $6$, $8$, $10$, $12$, $16$, $20$, $24$, $32$,
    $40$, $48$, $56$, $60$, $64$, $80$ and for path products (blue
    points) locations $U^*$ and $U^+$ for linear order $L=4$, $6$,
    $8$, $12$, $16$, $24$, $32$, $40$.}
  \label{fig:kdminloc4d}
\end{figure}

In Figure~\ref{fig:kdminloc4d} we show the location of the local
minima for both graph types. The location of the minima for the cycle
products are clearly approaching $U_c\approx 0.770527$, see
\cite{4dart}. However, the local minimum $U^*$ for the path products
does not appear to approach the same limit for our current set of
points.  We believe this is only due to the lattices being too small;
our current record size is only $L=40$ for the path products.  The
second minimum's location $U^+$ could very well be the same as for the
cycle products limit minimum location. An alternative possible
scenario is that the local minimum $U^*$ will indeed have a different
limit than $U_c$, but that this does not effect the behaviour of the
canonical quantities since their behaviour is guided by the global
minimum, the location of which appears to approach $U_c$.

The location $U^*$ of the minimum for the cycle products were set to
have the limit $c_0=U_c=0.770527$. This gave $c_1=4.14$ and
$\lambda_1=1.77$ when setting $c_2=0$ and fitting to the ten largest
lattice sizes. This looks very convincing. For the path products'
local minimum $U^*$ we received $c_0=0.7528$, i.e. a different limit,
and $c_1=23.1$, $\lambda_1=2.39$ when using $c_2=0$ and fitting to the
last five data points. The local minimum $U^+$, which is a global
minimum for $L\ge 24$, gave an acceptably good fit using $c_0=U_c$,
$c_1=0.788$, $\lambda_1=0.355$.

\section{5D-lattices}
\begin{figure}[!hbt]
  \begin{center}
    \includegraphics[width=0.49\textwidth]{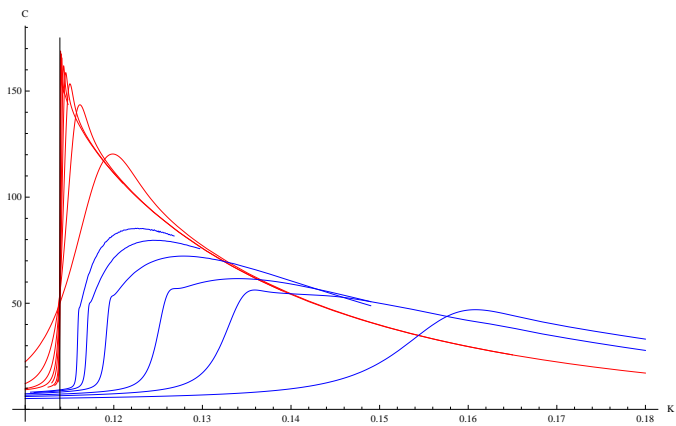}
  \end{center}
  \caption{(Colour online) Specific heat $\fC(K)$ for 5D cycle
    products (red) of linear order $L=4$, $6$, $8$, $10$, $12$, $16$,
    $20$, $24$ and for path products (blue) of linear order $L=4$,
    $6$, $8$, $12$, $16$, $20$. Black vertical line at $K_c\approx
    0.113914$.}
  \label{fig:c5d}
\end{figure}

For the 5-dimensional lattice the difference in behaviour between path
products and cycle products becomes even more pronounced.  Let us
begin with the specific heat as usual. As Figure~\ref{fig:c5d} shows,
the behaviour is similar to that of the 4-dimensional
lattice. 

\begin{figure}[!hbt]
  \begin{center}
    \includegraphics[width=0.49\textwidth]{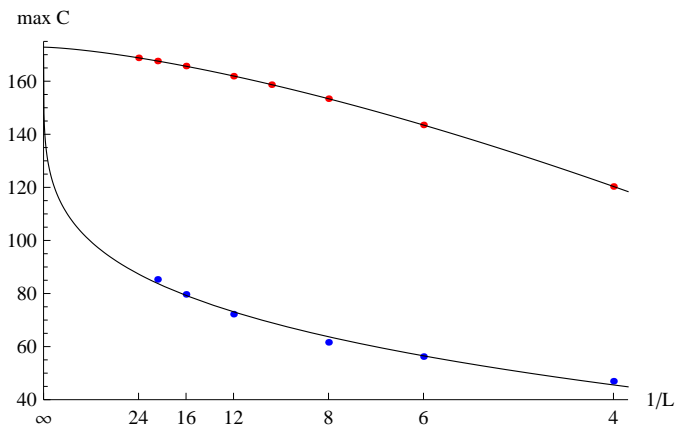}
  \end{center}
  \caption{(Colour online) Maximum specific heat $\max \fC(K)$ for 5D
    cycle products (red) of linear order $L=4$, $6$, $8$, $10$, $12$,
    $16$, $20$, $24$ and for path products (blue) of linear order
    $L=4$, $6$, $8$, $12$, $16$, $20$. 
  }
  \label{fig:cmaxval5d}
\end{figure}

For these lattices the specific heat is rigorously known
\cite{sokal:79} to be bounded. Using our
scaling formula (\ref{scale}) with $c_2=0$ we received $c_0=172.8$,
$c_1=-383.1$ and $\lambda_1 = 1.434$ giving an excellent fit over all
the lattices. This is demonstrated by the red points and their fitted
curve in Figure~\ref{fig:cmaxval5d}. Ideally we should also receive
the same asymptotic value for the path products though the finite size
scaling may be quite different. In the figure we have forced $c_0$ to
the same value, $172.8$, as for the cycle products. This gave
$c_1=-173.1$ and $\lambda_1=0.2222$. The fit does not look too
strained, but forcing the value of $c_0$ was necessary for this.

\begin{figure}[!hbt]
  \begin{center}
    \includegraphics[width=0.49\textwidth]{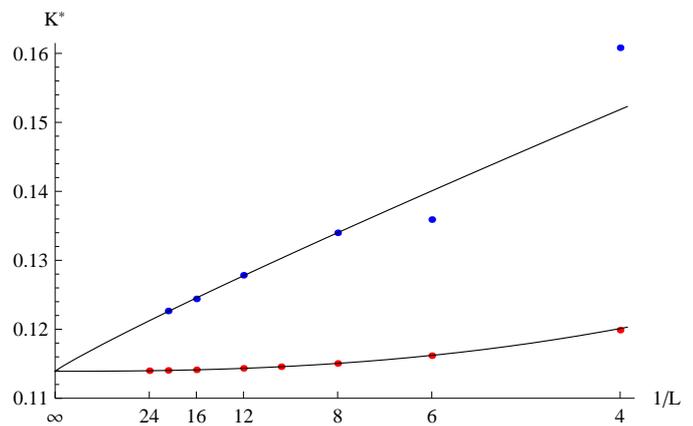}
  \end{center}
  \caption{(Colour online) Location $K^*$ of the maximum specific heat
    $\fC(K)$ for 5D cycle products (red) of linear order $L=4$, $6$,
    $8$, $10$, $12$, $16$, $20$, $24$ and for path products (blue) of
    linear order $L=4$, $6$, $8$, $12$, $16$, $20$.}
  \label{fig:cmaxloc5d}
\end{figure}

The locations of the maximum specific heat behave in the typical
smooth fashion for the cycle products but the path products have the
peculiar behaviour found in the 4D-lattice. Here, for $L\ge 8$ the
maximum has jumped farther out from $K_c$. This shows clearly in
Figure~\ref{fig:cmaxloc5d} where we plot the location of the maxima.

For the fitted curves we have used $c_0=K_c=0.113914$, see
\cite{luijten:99}.  Our best fit gave $c_1=0.183$ and $\lambda_1=2.44$
for the cycle products, but asymptotically we should have
$\lambda_1=5/2$. For the path products we found $c_1=0.135$ and
$\lambda_1=0.918$, based on the points for $L\ge 8$. The exponents
thus
differ considerably so there could be a different scaling rule in play
here.

\begin{figure}[!hbt]
  \begin{center}
    \includegraphics[width=0.49\textwidth]{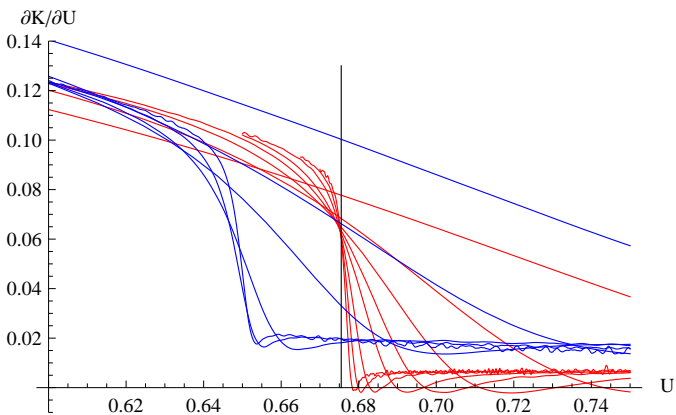}
  \end{center}
  \caption{(Colour online) $\partial K/ \partial U$ for 5D cycle
    products (red) of linear order $L=4$, $6$, $8$, $10$, $12$, $16$,
    $20$, $24$ and for path products (blue) of linear order $L=4$,
    $6$, $8$, $12$, $16$, $20$. Black line at $U_c\approx
    0.67549$.}
  \label{fig:kd5d}
\end{figure}

\begin{figure}[!hbt]
  \begin{center}
    \includegraphics[width=0.49\textwidth]{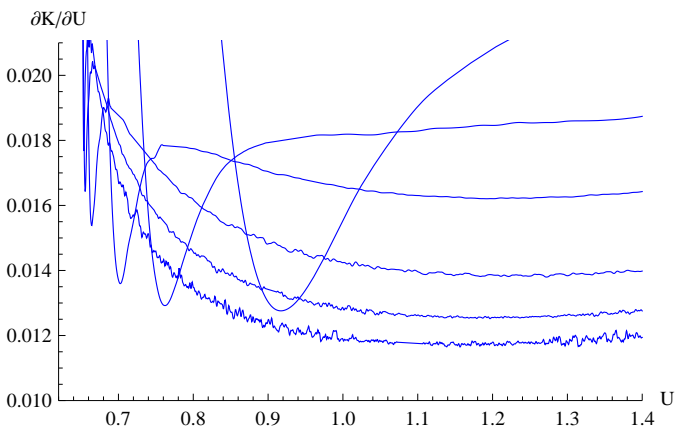}
  \end{center}
  \caption{(Colour online) $\partial K/ \partial U$ for 5D path
    products of linear order $L=4$, $6$, $8$, $12$, $16$, $20$.}
  \label{fig:kdp5d}
\end{figure}

Moving on to the microcanonical side we show in Figure~\ref{fig:kd5d}
the $\partial K/\partial U$-curves for the two types of lattices.  As
for the 4D-lattices we again see a huge qualitative difference between
them.  To begin with, for the path products we see a second local
minimum in the curve, located far away from $U_c$, which becomes the
global minimum for $L\ge 12$, see Figure~\ref{fig:kdp5d}.  Also, the
minima of the cycle products are all negative, whereas for the path
products they are all positive. This is a striking difference, since
a negative minimum is normally seen as an indication of a first-order
phase transition, and as we shall see later there is indeed a
first-order-like behaviour at the critical point for the cycle
products.

\begin{figure}[!hbt]
  \begin{center}
    \includegraphics[width=0.49\textwidth]{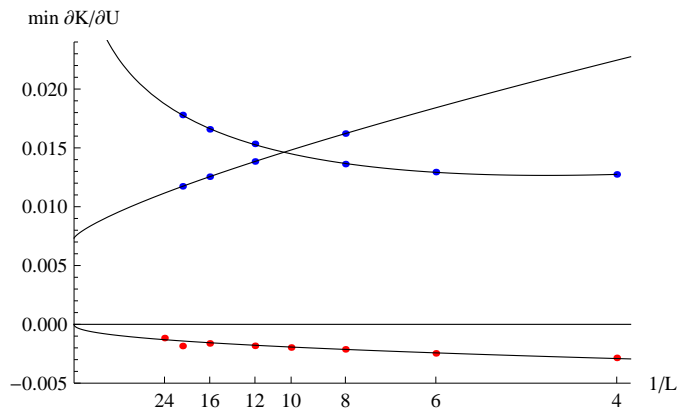}
  \end{center}
  \caption{(Colour online) Minimum value of $\partial K/ \partial U$
    for 5D cycle products (red points) of linear order $L=4$, $6$,
    $8$, $10$, $12$, $16$, $20$, $24$ and for path products (two sets
    of blue points) of linear order $L=4$, $6$, $8$, $12$, $16$,
    $20$. The blue points converging to $0.06$ (outside the plot) are
    the minima at $U^*$ and the blue points converging to $0.007$ are
    the minima at $U^+$.}
  \label{fig:kdminval5d}
\end{figure}

As for the values of the minima we turn our eyes to
Figure~\ref{fig:kdminval5d}. The fitted curve for the cycle products
has been forced to $c_0=0$, which may or may not be correct. Then
$c_1$ took the best-fit value $-0.0054$ and $\lambda_1$ chose $0.447$,
when setting also $c_2=0$. The point for $L=20$ was left out of the
fitting process since it clearly deviates from the others in its
behaviour. 

For the path products we now have two sets of minima to look at, just
as for the 4D-case. We run into trouble for the minima nearest $U_c$
though.  Fitting (\ref{scale}) with only one exponent gave negative
$\lambda_1$. This would imply that the minimum goes to infinity, which
in its turn implies that the specific heat would go to zero. This is
clearly an inconsistent behaviour due a too simple fit.
 Using two exponents instead, a best-fit gave
$c_0=0.062$, $c_1=-0.071$, $c_2=0.042$, $\lambda_1=0.151$ and
$\lambda_2=1.12$. The fit is very good. However, it should be kept in
mind that this means we have fitted 5 parameters to 6 points, so the
value of $c_0$ (as the other parameters) is likely to change upon
receiving further data for larger lattices. What seems clear is that
this minima has an increasing trend with increasing $L$. 

Considering instead the minima farther out from $U_c$ we see in the
figure that it becomes a global minimum for $L\ge 12$. Also, this
local minimum only exists for $L\ge 8$. The fitted curve, setting
$c_2=0$, has the parameters $c_0=0.0073$, $c_1=0.044$,
$\lambda_1=0.77$,
which gives a convincing fit but it relies on only 4 points. 

\begin{figure}[!hbt]
  \begin{center}
    \includegraphics[width=0.49\textwidth]{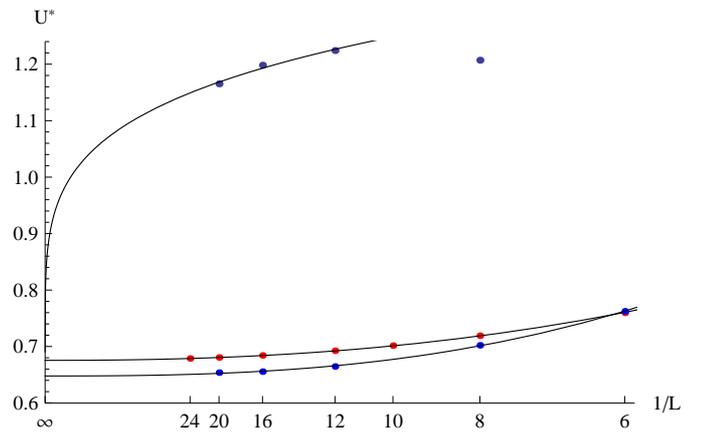}
  \end{center}
  \caption{(Colour online) Location $U^*$ of the minimum value of
    $\partial K/ \partial U$ for 5D cycle products (red points) of
    linear order $L=4$, $6$, $8$, $10$, $12$, $16$, $20$, $24$ and for
    path products ($U^*$ and $U^+$) of linear order $L=4$, $6$, $8$,
    $12$, $16$, $20$. The points at $U^+$ are above those at $U^*$.}
  \label{fig:kdminloc5d}
\end{figure}

Figure~\ref{fig:kdminloc5d} displays the location $U^*$ of the minimum
$\partial K/\partial U$-values. For the path products we thus see two
sets of these, $U^*$ and $U^+$. The curve fitted to the red points,
i.e. the cycle products, has been forced to $c_0=U_c=0.67549$ and this
gave $c_1=5.67$ and $\lambda_1=2.34$. Setting $c_0$ to $U_c$ for the
path products' minima at $U^*$ gave the curve a very strained look. A
free fit instead gave $c_0=0.6478$, $c_1=13.6$ and
$\lambda_1=2.67$. Receiving different limit values (which we also did
for the 4D-lattice) is a rather unsatisfactory outcome, but the data
points are very nicely fitted to this curve, and looks good even if we
were to zoom in on these points only. More likely, the lattices are
far too small to have received a correct asymptotic
behaviour. However, we should not rule out the possibility of a second
asymptotic critical point. After all, what matters to the model is
that the global minimum goes to the same value as for the cycle
products. This minimum's location $U^+$ is also seen in the figure but
it is very far from $U_c$. The curve through the last three points
only, has a forced $c_0=U_c=0.67549$ and this gave $c_1=0.946$ and
$\lambda_1=0.218$.

\begin{figure}[!hbt]
  \begin{center}
    \includegraphics[width=0.49\textwidth]{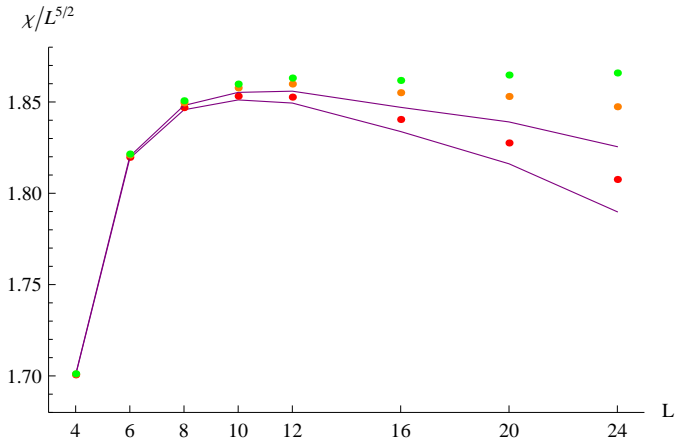}
  \end{center}
  \caption{(Colour online)$\chi/L^{5/2}$ vs $L$ for 5D cycle products
    of linear order $L=4,6,8,10,12,16,20,24$ at $K=0.1139139$ (red
    points), $K=0.1139150$ (orange points), $K=0.1139155$ (green
    points).  The purple lines correspond to the upper and lower ends
    of the estimate $K_c=0.1139139(5)$.}
  \label{fig:chic-5d}
\end{figure}

One focus of the earlier mentioned debates regarding the 5-dimensional
Ising model \cite{CD:00,LBB:99} was the scaling with $L$ of the
susceptibility at $K_c$. Here the main term in the scaling has been
predicted \cite{PhysRevB.31.1498,PF:85} to be $L^{5/2}$ for cyclic
boundary conditions, whereas an ordinary scaling would have been
proportional to $L^{1/\nu}=L^2$. In addition \cite{CD:00} predicted
that there should be additional terms leading to a maximum in
$\chi/L^{5/2}$ at $L=9$ after which it should decrease to a non-zero
limit, a prediction which was criticized in \cite{LBB:99}.  Given our
current, larger, data set we have made a brief return to this
question, for both boundary conditions.  In Figure \ref{fig:chic-5d}
we see $\chi/L^{5/2}$ for several values of $K$ close to $K_c$. That
the quotient is converging to a finite non-zero limit seems quite
plausible for a range of possible $K_c$, but with the limited system
sizes available to us it is hard to say what the limit should be.  In
\cite{CD:00,LBB:99} data for $K=0.1139150$ were used, corresponding to
the orange points in our figure. However, this value is larger than our
best estimate and as the figure shows the quotient is extremely
sensitive to even small changes in the value of $K_c$.  The data does
support a maximum in the quotient, but a location at $L=12$ seems more
likely than the $L=9$ predicted in \cite{CD:00}.

\begin{figure}[!hbt]
  \begin{center}
    \includegraphics[width=0.49\textwidth]{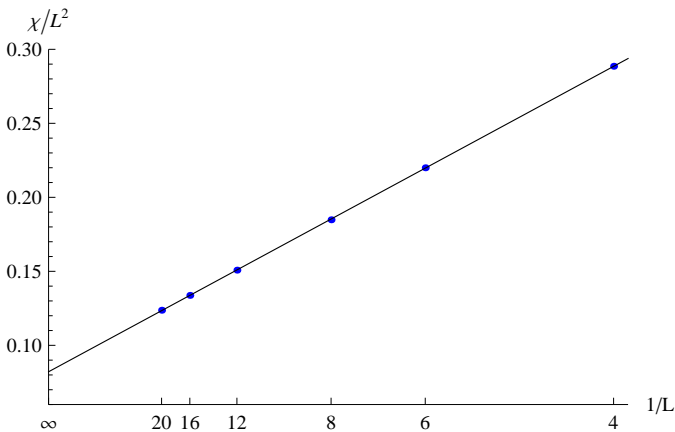}
  \end{center}
  \caption{(Colour online)$\chi/L^2$ vs $1/L$ for 5D path products of
    linear order $L=4,6,8,12,16,20$ at $K=0.1139139$ (blue
    points). The line has slope $0.825$ and the asymptotic value is
    $0.0821$.}
  \label{fig:chip-5d}
\end{figure}

In Figure \ref{fig:chip-5d} we plot the quotient $\chi/L^2$,
corresponding to $\chi/L^{1/\nu}$ , for open boundary conditions. We
see an excellent agreement over the full range of system sizes. We
thus conclude that for open boundary conditions the dominant term in
the scaling is $L^2$ and that any correction terms are significantly
smaller. Experiments also showed that this quotient was much less
sensitive to the value of $K_c$ than for cyclic boundaries.

\subsection{The nature of the phase transition for the cycle products}

The 5-dimensional cycle products are quite exceptional since the
$\partial K/\partial U$ become negative.  This is then obviously
reflected in the function $K(U)$ by having a local maximum and a local
minimum. This, in turn, means that the distribution of energies becomes
bimodal for a range of temperatures, see~\cite{ellis:04,reconart} for
discussions from the mathematical and computational perspectives
respectively.  A bimodal energy distribution is normally the clearest
sign of first-order, or discontinuous, phase transition, with a latent
heat given by the distance between the two maxima of the energy
distrubution. For $D>4$ the Ising model has been rigorously proven
\cite{aizenman:81,aizenman:82} to have a second-order phase transition
with critical exponents given by the mean field model.  Hence we here
see a clear difference between the asymptotic and finite size
behaviour, which we will now examine closer.

\begin{figure}[!hbt]
  \begin{center}
    \includegraphics[width=0.49\textwidth]{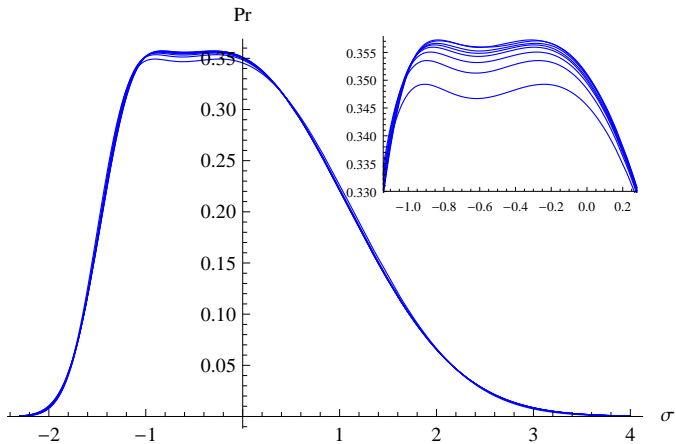}
  \end{center}
  \caption{(Colour online) Normalised distribution of energies at the
    temperature where the two peaks of the density function are equal,
    for 5D cycle products of linear order $L=4$, $6$, $8$, $10$, $12$,
    $16$, $20$, $24$ (blue).  The inset picture shows a zoomed-in
    version.
  }
  \label{fig:dist5d0}
\end{figure}

In Figure~\ref{fig:dist5d0} we show the normalised distributions at
the temperature where the two distribution peaks are equal and the
inset picture shows a zoomed-in plot clearly demonstrating that the
distributions are indeed bimodal. Denoting the location of the local
maxima of this bimodal distribution as $U_1$ and $U_2$ their
difference goes to zero as roughly $U_2-U_1 \sim 6.2/L^{5/2}$. The
right maximum $U_2$ behaves as roughly $U_c+13/L^{5/2}$.  The
temperature where the maxima of the distribution are of equal height
scales as $K_c+0.12/L^{5/2}$.

These results mean that for finite $L$ we see an effective latent heat
with size proportional to $L^{-5/2}$ and, as for other models with a
first order phase transition in their thermodynamic limit, the finite
size model demonstrates a bistable behaviour at $K_c$. This behaviour
is noticeable in the simulations, where the system will spend a long
time at one maximum and then rapidly switch to the other.

Given the scaling of $U_2-U_1$ we find it quite possible that the
model will display this bimodal distribution for all finite $L$ and
only reach the pure mean field behaviour in the thermodynamic limit
$L\rightarrow \infty$.

\begin{figure}[!hbt]
  \begin{center}
    \includegraphics[width=0.49\textwidth]{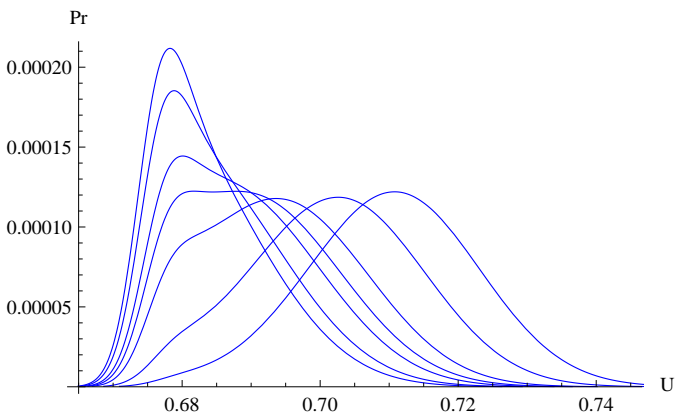}
  \end{center}
  \caption{(Colour online) Density functions for the 5D cycle products
    $L=16$ near $K_c$ at $K=0.11396$, $0.11398$, $0.11401$,
    $0.114027$, $0.11405$, $0.11410$, $0.11415$ (left to right).}
  \label{fig:dist5d1}
\end{figure}

The density function of the energy distribution for the cycle products
undergoes some dramatic shape-shifting near $K_c$. It starts as a
gaussian function, then becomes heavily skewed to the left, then
bimodal, then skewed to the right, and finally it becomes gaussian
again. In Figure~\ref{fig:dist5d1} we show this as a gallery of
distributions for $L=16$ near $K_c$.

\begin{figure}[!hbt]
  \begin{center}
    \includegraphics[width=0.49\textwidth]{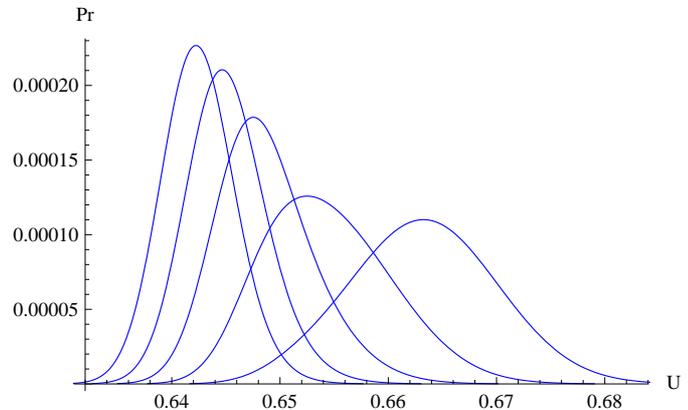}
  \end{center}
  \caption{(Colour online) Density functions for the 5D path products
    $L=16$ near $K_c$ at $K=0.1164$, $0.1166$, $0.1168$, $0.1170$,
    $0.1172$ (left to right).}
  \label{fig:dist5d2}
\end{figure}

The path products on the other hand, for which $\partial K/\partial U$
stays positive, clearly can not have a bimodal
distribution. Consequently, the density function has a rather
non-spectacular behaviour near $K_c$ as Figure~\ref{fig:dist5d2}
shows. Even at the temperature where the specific heat is at its
maximum the distribution stays quite gaussian. However, at the
temperature range where specific heat grows most quickly, there is
some typical behaviour. For example, the usual left-to-right skewness
is found, but it is considerably less pronounced than for the cycle
products.

\section{Conclusions}
As we have seen, the finite size behaviour of the Ising model becomes
significantly more complex as we pass the critical dimension $D=4$.
In the case of 4 dimensions we have found that an additional maximum
in the specific heat appears for a certain range of sizes, but only for
open boundary conditions. We also found that both the cyclic and
open cases seem to favour a bounded specific heat, with a value which
they approach from below and above respectively, thereby contradicting
a prediction from renormalization theory \cite{ptcp}.

Some of the peculiarities of the 4-dimensional cases appears in the
5-dimensional model as well, but here the most striking feature is the
quasi first order phase transition for cyclic boundary conditions. The
presence of this, for the Ising model, unusual phase transition type,
has numerous implications for Monte Carlo experiment and their
analysis.

\begin{enumerate}
	\item The model is highly sensitive to small changes in
          $K$. In Figure \ref{fig:dist5d1} the two curves closest to
          the bimodal one corresponds to the values $K=0.11401$ and
          $K=0.11405$. Here changes in the 5th decimal gives a
          significant change in the expected internal energy of the
          model.
	
	\item Simulations at values $K$ at the balanced bimodal value
          will display a meta-stable behaviour, in which the models
          will stay for some time at one maximum and then rapidly
          change to the other.
	
	\item Unlike what was initially hoped, the Wolff-cluster
          algorithm \cite{wolff:89} suffers from critical slowing down
          at first-order phase transitions, as was rigorously proven
          in \cite{slow}. This means that, just as for the Metropolis
          algorithm, additional care must be taken in the study of
          this model in order to make sure that the generated data
          does not suffer from unwanted correlations
	
	\item If only the moments of the energy distribution are
          studied at a given temperature the clearly non-gaussian form
          of the energy distribution can potentially lead to incorrect
          conclusions regarding higher moments based on the low order
          moments. There will e.g. be significantly non-zero odd
          moments for a longer interval of $K$ near $K_c$.
		
\end{enumerate}

As we mentioned in the introduction the finite size behaviour of the
simple percolation model has also been studied in higher
dimensions. In \cite{aizenman:97} Aizenman found that the size of the
largest cluster at $p_c$ scales, at most, as $L^4$ and conjectured
that for cyclic boundary condition the scaling should be of order
$L^{2d/3}$ instead. Recently this conjecture was proven in
\cite{HH:07}, apart from a logarithmic factor which was removed in
a follow up paper by the same authors \cite{HH:09}.

The general assumption in simulation physics is that as the system
grows larger the effect of the boundary should diminish and the model
converge to the same thermodynamic limit independently of the choice
of boundary condition and the rough shape of the finite
region. Following the seminal work of van Hove \cite{vH:49} we know
that for a large class of models this will be the case, but it is
worth to notice that van Hove's proof does in fact not apply to a
sequence of larger and larger lattices with cyclic boundary
conditions, as the proof assumes that the edge and vertices of the
smaller lattices are subsets of those of the larger. However, for $K$
well away from $K_c$ we see no reason to expect anything else.
Nonetheless, as the percolation results indicate, it is quite
plausible that when we stay at, or near, $K_c$ we will see drastically
different scalings depending on the choice of boundary conditions.

Generalizing the conjecture of Aizenman \cite{aizenman:97} we pose the
following conjecture for the $q\geq 1$ random cluster model
\cite{FK:72}
\begin{conjecture}
  There exists a constant $c(q)$ and a dimenison $D(q)$ such that for
  the random cluster model with $q\geq 1$ and $d\geq D(q)$, the
  largest cluster in the $d$-dimensional cycle product with side $L$
  contains $L^{c(q)d}$ vertices at $p_c(q)$.
\end{conjecture}
The case $q=1$ corresponds to percolation and $q=2$ to the Ising
model. We intend to return to this conjecture in an upcoming paper.

For models more complicated than the Ising model, such as those from
QCD \cite{CCC:98,Stevenson:05,BN:06}, quantitative differences between
different boundary conditions have certainly been expected. The
unexpected appearance of \emph{qualitative} differences in the nature
of the finite-size phase transition, such as what we have seen for
$D=5$, poses a more serious problem. Close to a phase transition point
different boundary conditions can potentially lead to quite distinct
asymptotics, thereby making it necessary to study more than one kind
of boundary in order to help pinpoint the correct infinite system
asymptotics.

On a more general note we believe that the sensitivity to $K$
displayed in Figure \ref{fig:chic-5d} clearly demonstrates why single
value simulations of models no longer suffice for studying models at
the level of accuracy needed in modern physics. Instead methods
allowing a continuous variation of the system parameters after the
simulation, such that of \cite{sampart}, are needed. The authors of
\cite{LBB:99} stated that "for a full understanding of the problem a
variation of parameters over a broad range is desirable" and looking
at our current findings we most emphatically agree.


\section*{Acknowledgements}
This research was conducted using the resources of High Performance
Computing Center North (HPC2N).


\end{document}